\pdfoutput=1
\documentclass[prl,aps,preprintnumbers,floats,floatfix,superscriptaddress,preprintnumbers,showpacs,eqsecnum,nofootinbib,notitlepage,twocolumn]{revtex4-2}
\usepackage[utf8]{inputenc}
\usepackage{latexsym,array,theorem,mathrsfs,bm,float}
\usepackage{psfrag}
\usepackage{amsfonts,amsmath,amssymb,afterpage,epsfig,color,graphicx,tabularx,here,multirow}

\newcommand{\Mpl}{M_{\rm Pl}}

\newcommand{\wh}{\omega_h}
\newcommand{\wt}{\omega_t}

\newcommand{\ncpl}{{\alpha}}

\newcommand{\D}{{\rm d}}

\begin{document}
\baselineskip=12pt

\title{Primordial-tensor-induced stochastic gravitational waves}
\author{Mohammad Ali Gorji}
\email{gorji@icc.ub.edu}
\affiliation{Departament de F\'{i}sica Qu\`{a}ntica i Astrof\'{i}sica, Institut de Ci\`{e}ncies del Cosmos, Universitat de Barcelona, Mart\'{i} i Franqu\`{e}s 1, 08028 Barcelona, Spain }
\affiliation{Center for Gravitational Physics and Quantum Information, Yukawa Institute for Theoretical Physics, Kyoto University, Kyoto 606-8502, Japan}

\author{Misao Sasaki}
\email{misao.sasaki@ipmu.jp}
\affiliation{Kavli Institute for the Physics and Mathematics of the Universe (WPI), The University of Tokyo, 277-8583, Chiba, Japan}
\affiliation{Center for Gravitational Physics and Quantum Information, Yukawa Institute for Theoretical Physics, Kyoto University, Kyoto 606-8502, Japan}
\affiliation{Leung Center for Cosmology and Particle Astrophysics,
National Taiwan University, Taipei 10617, Taiwan }

\date{\today}

\begin{center}
\begin{abstract}
Cosmological stochastic gravitational waves (GWs) induced by a spectator field are usually expected to have an amplitude very small compared with those generated by the curvature perturbation, or equivalently by a field dominating the universe.
On the contrary to this expectation, we show that a spectator field that provides a tensor perturbation, on top of the metric tensor perturbation, can generate a significant amount of GWs. The amplitude and frequency of the generated GWs may lie within the sensitivity range of future GW detectors. In particular, if the sound velocities of the two tensor perturbations coincide,
the induced GW amplitude may become very large due to resonance by forced oscillation, even in
the limit of small coupling between them.
A distinct feature of this scenario is that, since tensor modes can hardly lead to the formation of primordial black holes (PBHs), we expect no presence of PBHs, 
in contrast to the usual scalar-induced case, in which the detection of strong enough induced GWs suggests the existence of PBHs.
\end{abstract}
\end{center}

\maketitle

{\bf Introduction.} 
The new generations of GW detectors like LIGO/VIRGO/KAGRA~\cite{KAGRA:2021kbb}, 
ET~\cite{Punturo:2010zz}, 
DECIGO~\cite{Seto_2001,Yagi_2011,Kawamura:2020pcg}, 
LISA~\cite{LISA:2017pwj,Barausse_2020,LISACosmologyWorkingGroup:2022jok}, PTA~\cite{Lentati:2015qwp,NANOGrav:2020bcs}, 
Taiji~\cite{Ruan_2020}, and TianQin~\cite{TianQin:2015yph} 
can probe high frequency $k\sim10^{7}-10^{18}\,\mbox{Mpc}^{-1}$ inflationary perturbations that are not accessible at the cosmic microwave background (CMB) scale $k_{CMB}\sim0.05\,\mbox{Mpc}^{-1}$. 
In recent years, this opportunity is seized to probe the physics at late stages of inflation. In particular, a lot of attention has been paid to the roles of spectator fields during inflation.

Commonly, however, a spectator field can produce only a small amount of GWs 
because of its small contribution to the energy-momentum tensor.
While this is the case for typical {\it scalar} and {\it vector} fields,
a spectator field that provides {\it tensorial} perturbations on top of the metric tensor perturbation can make significant contributions
to the GWs \cite{Caldwell:2016sut,Thorne:2017jft,Iacconi:2019vgc,Iacconi:2020yxn,BeltranJimenez:2019xxx,LISACosmologyWorkingGroup:2019mwx}. This type of spectator field shows up, i.e., in bi-gravity theories \cite{deRham:2010kj,LISACosmologyWorkingGroup:2019mwx}, when a non-Abelian gauge field acquires homogeneous and isotropic vev \cite{Maleknejad:2011jw}, in modified gravity theories with dynamical torsion \cite{Aoki:2019snr}, and when there will be a spin-2 (or higher spin) field \cite{Bordin:2018pca}. 

To see this fact, let us look at the equations of motion for the transverse-traceless metric tensor perturbation $h_{ij}$, which characterizes GWs, in a spatially flat Friedmann‐Lema\^itre‐Robertson-Walker (FLRW) background,
\begin{equation}\label{EoM-iGWs}
h''_{ij} + 2 \frac{a'}{a} h'_{ij} - \partial^2 h_{ij} = S^{\rm TT}_{ij} \,,
\end{equation} 
where a prime denotes derivative with respect to the conformal time $\tau$, 
$a$ is the scale factor and $S^{\rm TT}_{ij}$ is the transverse-traceless source,
which may be schematically expressed in an expansion form,
\begin{equation}\label{source-expansion}
S^{\rm TT}_{ij} = {\cal O}(\epsilon_T) + {\cal O}(\epsilon_S^2) + {\cal O}(\epsilon_V^2) +  {\cal O}(\epsilon_T^2) 
+ \cdots
\,,
\end{equation} 
where $\epsilon_S$, $\epsilon_V$, and $\epsilon_T$ represent the amplitudes of scalar, vector, and tensor perturbations.
In the absence of any extra tensor modes, the curvature perturbation,
corresponding to ${\cal O}(\epsilon_S^2)$ in the above,
gives the dominant contribution \cite{Ananda:2006af,Baumann:2007zm,Alabidi:2012ex,Espinosa:2018eve,Kohri:2018awv} leading to the usual scenario of the {\it secondary} scalar-induced GWs \cite{Domenech:2021ztg}. 
However, if there exist extra tensor modes, they can contribute already at linear order. In this paper, we consider a setup that provides such tensor modes,
and compute the corresponding {\it primary} tensor-induced GWs.\\


{\bf The model.}
We exploit an effective field theory approach which captures essential and universal features of transverse-traceless extra tensor modes $t_{ij}$.
The quadratic action for $t_{ij}$, which is minimally coupled to gravity,
is given by ($\hbar=1=c$) \cite{Bordin:2018pca}
\begin{align}
S &= \frac{1}{2} \int \D^3x\, \D\tau\, a^2 \left[ 
\left( {\gamma}'_{ij} \right)^2 
- \left( \partial_i{\gamma}_{jk} \right)^2
\right]
\nonumber \\ \nonumber
&+
\frac{1}{2} \int \D^3x\, \D\tau\, a^2 f^2 \left[ 
\left( {t}'_{ij} \right)^2 
- c_t^2 \left( \partial_i{t}_{jk} \right)^2
\right]
\\ \label{action}
&+ 
\int \D^3x\, \D\tau\, a^3 \ncpl H \left[ t^{ij} \gamma'_{ij} \right] \,,
\end{align}
where $\gamma_{ij}=\Mpl{h}_{ij}/2$ ($\Mpl=1/\sqrt{8\pi{G}}$) represents the standard metric tensor perturbation, $H=a'/a^2$ is the Hubble parameter, and $c_t$, $f$, $\ncpl$ are functions of time. For the sake of simplicity, we assume $t_{ij}$ is massless.\footnote{Indeed, action \eqref{action} represents massless limit of a massive spin-two field which has 5 degrees of freedom: 1 helicity-zero scalar mode, 2 helicity-one divergenceless vector modes and 2 helicity-two transverse-traceless tensor modes $t_{ij}$. As shown in \eqref{source-expansion}, assuming the same initial amplitudes for all 5 modes, $t_{ij}$ dominates at linear order.} Note that there can be other types of linear interactions between $\gamma_{ij}$ and $t_{ij}$ (see \cite{BeltranJimenez:2019xxx}) while the $\ncpl$-term in \eqref{action} is the most relevant one for our purpose. 

The equations of motion in Fourier space, $X_{ij}(\tau,{\bf k}) = \sum_{\lambda=+,\times} e^{\lambda}_{ij}(\hat{\bf k}) X^{\lambda}_{\bf k}(\tau)$
($X=\gamma, t$), where $e_{ij}^\lambda(\hat{\bf k})$ is the polarization tensor \cite{Caprini:2018mtu}, are 
\begin{align}\label{EoM-h}
&\gamma''^\lambda_{\bf k} + 2 \frac{a'}{a} \gamma'^\lambda_{\bf k} + k^2 {\gamma}^\lambda_{\bf k}
= - \ncpl \frac{a'}{a}
\left[
t'^{\lambda}_{\bf k} + \frac{(\ncpl a a')'}{\ncpl a a'} {t}^{\lambda}_{\bf k} 
\right] \,,
\\
\label{EoM-t}
&t''^\lambda_{\bf k} + 2 \frac{(af)'}{af} t'^\lambda_{\bf k} + c_t^2k^2 {t}^\lambda_{\bf k}
= 
\frac{\ncpl}{f^2} \frac{a'}{a} \gamma'^{\lambda}_{\bf k} \,.
\end{align}


{\bf Tensor-induced GWs.}
We consider the scenario in which $\ncpl$ vanishes during inflation and gets a nonzero value during radiation dominance where $a\propto\tau$,
\begin{align}\label{int-def}
&\ncpl_{\rm inf} = 0 \,,
&\ncpl_{\rm r} \neq 0 \,.
\end{align}
The above condition is only an assumption to make the setup simple. The case $\ncpl_{\rm inf} \neq 0$ is an interesting possibility in which one deals with GWs production at different scales during and after inflation. However, for our purpose in this letter, the simple subset of the model given by \eqref{int-def} is enough.

\begin{figure}[htbp!]
	\centering
	\includegraphics[width=1 \columnwidth]{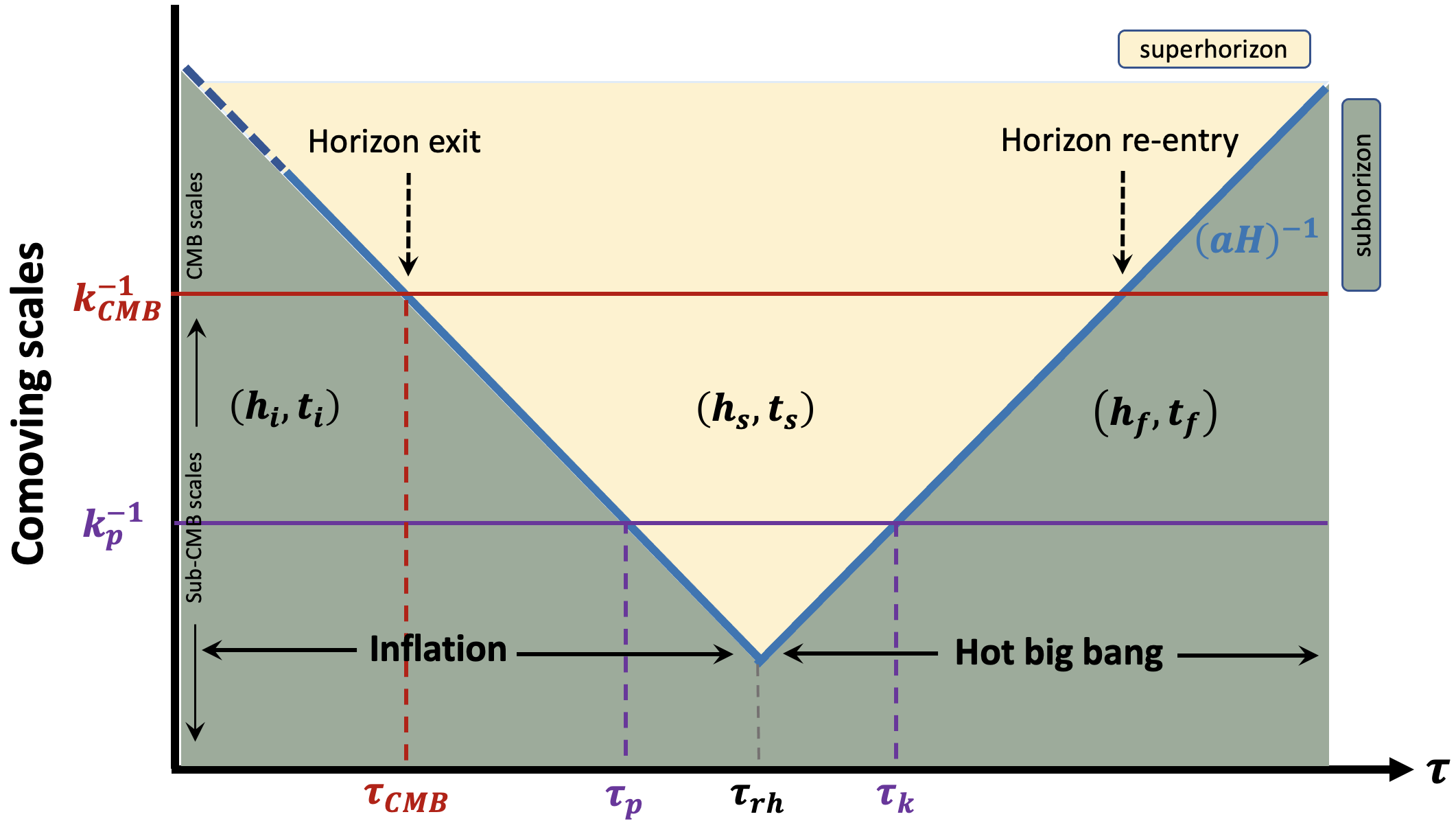}
	\caption{Tensor modes $t^\lambda_{\bf k}$ and $\gamma^\lambda_{\bf k}$ are decoupled $\ncpl_{\rm inf}=0$ during inflation. $t^\lambda_{\bf k}$ enhances at $\tau\sim\tau_p$ and sources $\gamma^\lambda_{\bf k}$ later during radiation dominance $\tau\sim\tau_f\gg\tau_k$ as $\ncpl_{\rm r}\neq0$. Thus, primary GWs with a peak at $k\sim{k}_p\gg k_{ CMB}$ will be generated from $t^\lambda_{\bf k}$.}
	\label{fig:1}
\end{figure}

We are interested in small scales around $k\sim k_p\gg k_{CMB}$ indicated in Fig.~\ref{fig:1}.
The spectral density fraction of high frequency GWs is
\begin{align}\label{GWs-sd}
&\Omega_{\rm GW}(k,\tau) = \frac{1}{12} \left(\frac{k}{aH}\right)^2
{{\cal P}_{h}(k,\tau)}
\,,
\end{align}
where the power spectrum is defined as ${\cal P}_{h}(k,\tau) = \sum_{\lambda}{\cal P}^\lambda_{h}(k,\tau)$ with $\langle {h}^\lambda_{\bf k}(\tau) {h}^{\ast{r}}_{{\bf q}}(\tau) \rangle 
= ({2\pi^2}/{k^3}) {\cal P}^\lambda_h(k,\tau) \delta^{\lambda{r}} \delta({\bf k}+{\bf q})$.

Initially, all the modes are on superhorizon scales. When the universe becomes radiation-dominated, $\ncpl_{\rm r}$ is turned on. 
Nevertheless, one can easily show that the effect of interaction is minimal while the modes are superhorizon.
Hence, we may assume that both $\gamma_{ij}$ and $t_{ij}$ are frozen until 
they re-enter the horizon. 
Namely, as we are interested in the tensor-induced GWs, we neglect the subdominant contribution from the vacuum fluctuation in comparison with the contribution from the enhanced amplitude of extra tensor mode and we set $\gamma_{ij}(\tau_{k,{\rm inf}})\approx0$ and
$t_{ij}=t_{ij}(\tau_{k,{\rm inf}})$ when $k\tau<1$, where $t_{ij}(\tau_{k,{\rm inf}})$ is the amplitude of the tensor mode when it left the horizon during inflation.
We then solve \eqref{EoM-h} and \eqref{EoM-t} inside the horizon during the
radiation dominance under the WKB approximation.

Deep inside the radiation-dominated $\tau\sim\tau_f$ with $k\tau\gg1$, Eqs. \eqref{EoM-h} and \eqref{EoM-t} simplify to
\begin{align}\label{EL-H}
&H''_{{\bf k}} + \wh^2 {H}_{{\bf k}}
= - \ncpl_{\rm r} {\cal H} 
T'_{{\bf k}}  \,; 
\quad 
\wh \equiv k \,,
\\
\label{EL-T}
&T''_{{\bf k}} +\wt^2 T_{{\bf k}}
= \ncpl_{\rm r} {\cal H} H'_{{\bf k}} \,;
\qquad
\wt \equiv c_tk \,,
\end{align}
where we dropped polarization indices for notational simplicity,
${\cal H}=a'/a$ is the conformal Hubble parameter, and
\begin{align}\label{H-T-def}
&{H}_{{\bf k}} = a\, {\gamma}_{\bf k} \,,
&&{T}_{{\bf k}} = a\, {t_{\bf k}} \,.
\end{align}
Substituting $H_{\bf k}$ and $T_{\bf k}$ given by
\begin{align}\label{H-T}
& {H}_{{\bf k}} = \frac{-i}{\wh} \left( {\mathsf c} \omega_+ X_{\bf k} +  {\mathsf d} \omega_- Y_{\bf k} \right) \,, 
&& T_{{\bf k}} = -{\mathsf d} X_{\bf k} + {\mathsf c} Y_{\bf k} \,,
\end{align}
in \eqref{EL-H} and \eqref{EL-T}, and setting ${\mathsf c}^2+ {\mathsf d}^2=1$,
we have
\begin{align}\label{Hamiltons-EoM}
&& X''_{\bf k} + \omega_+^2 X_{\bf k} =0 \,, 
&& Y''_{\bf k} + \omega_-^2 Y_{\bf k} =0 \,,
\end{align}
where
\begin{align}\label{omega-pm}
\omega_{\pm}^2 &= 
\frac{1}{2} \left[\wh^2+\wt^2 + \ncpl_{\rm r}^2 {\cal H}^2 \pm \sqrt{\Delta} \right] \,,\\
\label{alpha-def}
{\mathsf c} &\equiv \frac{1}{\sqrt{2}}
\left[
1+\frac{\wh^2-\wt^2- \ncpl_{\rm r}^2 {\cal H}^2}{\sqrt{\Delta}}
\right]^{1/2}
\,,\\ \label{beta-def}
 {\mathsf d} &\equiv 
\frac{1}{\sqrt{2}}
\left[
1-\frac{\wh^2-\wt^2- \ncpl_{\rm r}^2 {\cal H}^2}{\sqrt{\Delta}}
\right]^{1/2}
\,,
\end{align}
and
\begin{align}
&\Delta \equiv 
\left[\wh^2-\wt^2-\ncpl_{\rm r}^2 {\cal H}^2\right]^2 + 4 \ncpl_{\rm r}^2 {\cal H}^2 \wh^2 \,.
\end{align}
For the modes deep inside the horizon, the positive frequency WKB solutions of \eqref{Hamiltons-EoM} are
\begin{align}\label{X-Y}
&X_{\bf k} 
= i A \frac{\wh}{\omega_+} \frac{e^{-i\int^\tau \omega_+({\tilde \tau}) d{\tilde \tau}}}{\sqrt{2\omega_+}}\,,
&&Y_{\bf k} 
= B \frac{e^{-i\int^\tau \omega_-({\tilde \tau}) d{\tilde \tau}}}{\sqrt{2\omega_-}} \,,
\end{align}
where $A$ and $B$ are constants. Substituting \eqref{X-Y} in \eqref{H-T}, we find explicit WKB solutions of ${H}_{{\bf k}}$ and ${T}_{{\bf k}}$.

The constants $A$ and $B$ should be fixed by the initial conditions which are generated during inflation. Since we assume \eqref{int-def}, $\gamma_{\bf k}$ and $t_{\bf k}$ are decoupled during inflation. While $\gamma_{\bf k}$ has the conventional vacuum fluctuation amplitude ($\sim H_{\rm inf}$), which we ignore in this paper,
$t_{\bf k}$ can be enhanced due to a dip in $f$ \cite{Pi:2021dft} and/or $c_t$ \cite{Iacconi:2019vgc,Iacconi:2020yxn} around $\tau=\tau_p$ (see Fig.~\ref{fig:1}) without affecting $\gamma_{\bf k}$.
We assume $f=1$ and $c_t$ is constant or only slowly varying in time during radiation dominance.

As we mentioned before, $\gamma_{\bf k}\approx0$ and $t_{\bf k}\approx const.$ 
on superhorizon scales $k\tau\ll 1$.
We may assume this is a good approximation until the horizon 
crossing at $\tau=\tau_k$, and match them to subhorizon solutions given by \eqref{H-T}
at $\tau=\tau_k$ or $a_k=k/H_k$,
\begin{align}\label{H-HC}
& \frac{-i}{a_k\wh} \left[ \left( {\mathsf c} \omega_+ X_{\bf k} +  {\mathsf d} \omega_- Y_{\bf k} \right) \right]_{\tau=\tau_k} = \gamma_{\bf k}(\tau_k)=0\,, 
\\ \label{T-HC}
& \frac{1}{c_k} \left[- {\mathsf d} X_{\bf k} + {\mathsf c} Y_{\bf k} \right]_{\tau=\tau_k} 
=t_{\bf k}(\tau_k) \,,
\end{align}
where $t_{\bf k}(\tau_k)$ is given by the amplitude determined from inflation. 
Substituting \eqref{X-Y} in the above, we find $A$ and $B$ in terms of $t_{\bf k}(\tau_k)$.
Then the power spectra of $h_{ij}$ and $t_{ij}$ during radiation dominance are
given in terms of the power spectrum of $t_{ij}$ at $\tau=\tau_k$, 
${\cal P}_{t,k}$, as
\begin{align}\label{PSh-f}
{\cal P}_{h}(k,\tau) &= 
\left(\frac{a_k}{a}\right)^2 
K_h(k,\tau)
\ncpl_{\rm r}^2 c_t^2 \frac{4{\cal P}_{t,k}(k)}{\Mpl^2} \,,
\\ \label{PSt-f}
{\cal P}_{t}(k,\tau) &= 
\left(\frac{a_k}{a}\right)^2 
K_t(k,\tau)
{\cal P}_{t,k}(k) \,,
\end{align}
where we have defined the kernels,
\begin{align}\nonumber
&K_h(k,\tau) = \frac{1}{\ncpl_{\rm r}^2 c_t^2 \omega_h^2}
\left[
\frac{ {\mathsf c}_k  {\mathsf d}_k \omega_{-,k} \omega_{+,k}}{  {\mathsf d}_k^2 \omega_{-,k}+{\mathsf c}_k^2 \omega_{+,k} }
\right]^2 \\ \label{K-h-G}
&\times \left[
\frac{{\mathsf c}^2 \omega_{+,k}}{{\mathsf c}_k^2 \omega_{+}}
+
\frac{ {\mathsf d}^2 \omega_{-}}{ {\mathsf d}_k^2 \omega_{-,k}}
-
\frac{2{\mathsf c}  {\mathsf d}}{{\mathsf c}_k  {\mathsf d}_k} 
\sqrt{\frac{\omega_{-} \omega_{+,k}}{\omega_{-,k} \omega_{+}}}
\cos\theta
\right] , \\ \nonumber
&K_t(k,\tau) =\frac{1}{\omega_{-}^2}
\left[
\frac{{\mathsf c}_k {\mathsf d}_k\omega_{-,k} \omega_{+,k}}{ {\mathsf d}_k^2 \omega_{-,k}+{\mathsf c}_k^2 \omega_{+,k}}
\right]^2 
\\ \label{K-t-G}
&\times \left[\frac{{\mathsf c}^2}{ {\mathsf d}_k^2}
\frac{\omega_{-}}{ \omega_{-,k}}
+ \frac{ {\mathsf d}^2}{{\mathsf c}_k^2} \frac{ \omega_{-}^2\omega_{+,k}}{\omega_{+}^3}
+\frac{2{\mathsf c} {\mathsf d}}{{\mathsf c}_k  {\mathsf d}_k}
\frac{ \omega_{-}^{3/2}}{\omega_{+}^{3/2} } \sqrt{\frac{\omega_{+,k}}{\omega_{-,k}}} \cos\theta
\right] ,
\end{align}
with
\begin{align}\label{theta}
\theta \equiv \int d\tau \left[ (\omega_+-\omega_-) - (\omega_{+,k}-\omega_{-,k}) \right] \,,
\end{align}
and the subscript $k$ denoting the quantity evaluated at $\tau=\tau_k$. It is easy to check that $K_t(k,\tau_k)=1$ and $K_h(k,\tau_k)=0$ as desired by the initial conditions \eqref{H-HC} and \eqref{T-HC}.

Here, one interesting comment is in order. In principle, the coupled linear system \eqref{EoM-h} and \eqref{EoM-t} can always be diagonalized as \eqref{Hamiltons-EoM}. There are no oscillatory features in the spectrum of the decoupled modes $|X_{\bf k}|^2$ and $|Y_{\bf k}|^2$, as determined by the WKB solutions \eqref{X-Y}. However, as can be seen from the definition of the angle $\theta$ in \eqref{theta}, oscillatory features in \eqref{K-h-G} and \eqref{K-t-G} emerge as long as $\omega_+\neq\omega_-$. Since the coupled system \eqref{EoM-h} and \eqref{EoM-t} and \eqref{Hamiltons-EoM} are mathematically equivalent, one might question whether these oscillatory features are physical or not. The answer is that they are physical, as the observable quantity by the GW detectors is the amplitude of $h_{ij}$, which characterizes the fluctuation of spacetime. This process is very similar to the well-known phenomenon of neutrino oscillation when the system is decoupled in mass basis, while it is coupled in the flavor basis, which is the basis in which the detectors look.

\begin{figure}[htbp!]
	\centering
	\includegraphics[width=1 \columnwidth]{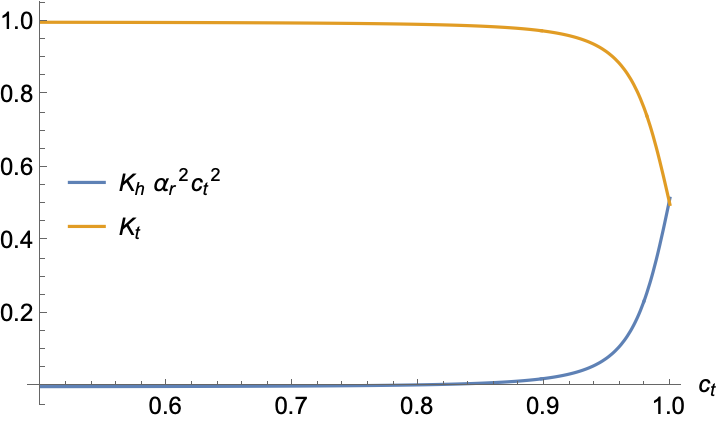}
	\caption{Kernels \eqref{K-h-G} and \eqref{K-t-G} as functions of sound speed $c_t$ for $\Delta{\cal N}\sim30$ around the LISA bound and $\ncpl_{\rm r}\sim\Delta{\cal N}^{-1}\sim3\times10^{-2}$. For $c_t$ close enough to unity, $K_h\ncpl_{\rm r}^2c_t^2={\cal O}(1)$ can be achieved. For larger values $\ncpl_{\rm r}\gtrsim\Delta{\cal N}^{-1}$, $K_h\ncpl_{\rm r}^2c_t^2={\cal O}(1)$ will be sooner achieved.}
	\label{fig:K-ct}
\end{figure}

\begin{figure}[htbp!]
	\centering
	\includegraphics[width=1 \columnwidth]{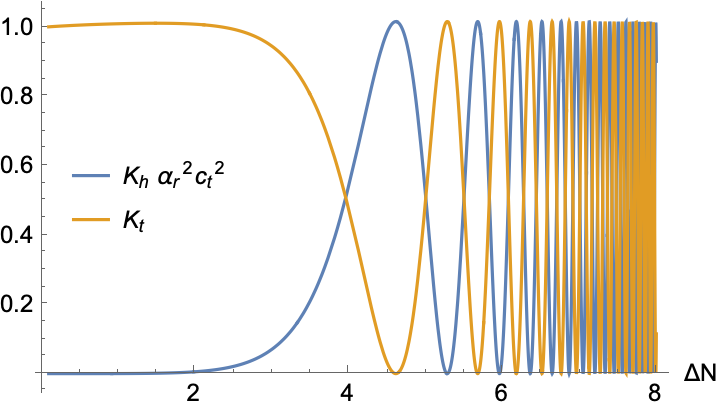}
	\caption{Kernels \eqref{K-h-G} and \eqref{K-t-G} as functions of number of e-folds for $c_t=1$ and $\ncpl_{\rm r}\sim3\times10^{-2}$. Due to the initial conditions \eqref{H-HC} and \eqref{T-HC}, $K_h=0$ and $K_t=1$ at the time of horizon crossing $\tau=\tau_k$ when $\Delta{\cal N}=0$. $K_h$ increases in time while $K_t$ decreases due to the GWs production from the extra tensor perturbation $t_{ij}$ through the mixing $\ncpl_{\rm r}\neq0$. At some point, $K_h$ and $K_t$ start to oscillate (as approximated in \eqref{K-h} and \eqref{K-t}) and, finally, their average amplitudes become of the order of unity $K_h\ncpl_{\rm r}^2c_t^2\sim1\sim{K}_t$. The large values $K_h\ncpl_{\rm r}^2c_t^2={\cal O}(1)$ can be achieved due to the resonance by forced oscillation which only happens for $c_t\sim1$. }
	\label{fig:K-N}
\end{figure}

In general, $K_h$ and $K_t$ have complicated forms. We thus try to understand their asymptotic behavior. Considering $\tau/\tau_k=a/a_k=\exp(\Delta{\cal N})$, 
where $\Delta{\cal N}=\ln(a/a_k)$ is the number of e-folds after the horizon crossing during the radiation dominance, $K_h$ and $K_t$ can be written as functions of $\Delta{\cal N}$ and two constant parameters $\ncpl_{\rm r}$ and $c_t$. Taking the limit $\ncpl_{\rm r}\ll1$ and look at two interesting different regimes, we find the following simple expressions
\begin{align}\label{K-h}
K_h(k,\tau) \ncpl_{\rm r}^2 &\approx \begin{cases}
\ncpl_{\rm r}^2 (1-c_t^2){}^{-2} & c_t < 1 \,,
\\
\sin^2(\ncpl_{\rm r}\Delta{\cal N}/2) &  c_t = 1 \,,
\end{cases}
\\ \label{K-t}
K_t(k,\tau) &\approx \begin{cases}
1 &c_t < 1 \,,
\\
\cos^2(\ncpl_{\rm r}\Delta{\cal N}/2) &c_t = 1 \,,
\end{cases}
\end{align} 
for the modes deep inside the radiation dominated era $k\tau\gg1$. 

The expression for $c_t<1$ in the first line of \eqref{K-h} is valid only up to $c_t$ for which its
value becomes equal to that of the second line. We have  illustrated the full expressions of $K_h$ and $K_t$, given by \eqref{K-h-G} and \eqref{K-t-G} respectively, as functions of $c_t$ in Fig.~\ref{fig:K-ct} for the fixed value of $\Delta{\cal N}\sim30$ which is around the LISA band. For $c_t\ll1$, we find $K_h\ncpl_{\rm r}^2\sim\ncpl_{\rm r}^2\ll1$ and $K_t\sim1$ while $K_h\ncpl_{\rm r}^2\sim1\sim{K}_t$ for $c_t\sim1$. The large values $K_h\ncpl_{\rm r}^2\sim1$ correspond to a very efficient GWs production which happens for $c_t=1$ especially when $\ncpl_{\rm r}\gtrsim\Delta{\cal N}^{-1}$. In Fig.~\ref{fig:K-N}, we have plotted the full expressions of $K_h$ and $K_t$, given by \eqref{K-h-G} and \eqref{K-t-G} respectively, for this interesting case. As it can be seen, due to the initial conditions \eqref{H-HC} and \eqref{T-HC}, $K_h=0$ and $K_t=1$ at the time of horizon crossing $\tau=\tau_k$ when $\Delta{\cal N}=0$. When the modes re-enter the horizon $\tau>\tau_k$, $K_h$ increases while $K_t$ decreases due to the GWs production from the enhanced extra tensor mode $t_{ij}$ through the mixing $\ncpl_{\rm r}\neq0$. For $\tau\gg\tau_k$, $K_h$ and $K_t$ start to oscillate (as approximated in \eqref{K-h} and \eqref{K-t}) and their average amplitudes become of the order of unity $K_h\ncpl_{\rm r}^2c_t^2\sim1\sim{K}_t$. The large values $K_h\ncpl_{\rm r}^2c_t^2={\cal O}(1)$ can be achieved due to the resonance by forced oscillation which only happens for $c_t\sim1$. As this case may have important observational consequences, we will discuss it in detail later.

Substituting \eqref{PSh-f} in \eqref{GWs-sd} yields the following expression for the energy density of GW in radiation dominated era
\begin{align}\label{GWs-sd-r}
\Omega_{{\rm GW,r}}(k)
= \frac{1}{3} {K_h(k,\tau)} \, 
\ncpl_{\rm r}^2 c_t^2 \frac{{\cal P}_{t,k}(k)}{\Mpl^2} \,.
\end{align}
Or, in terms of the fractional energy density of $t_{ij}$
\begin{align}\label{Omega-t}
\Omega_t(k,\tau) = \frac{1}{3}\left(\frac{c_tk}{aH}\right)^2 \frac{{ {\cal P}_{t}(k,\tau)}}{\Mpl^2} \,,
\end{align}
which is obtained in the similar way as \eqref{GWs-sd}, we find
\begin{align}\label{Omega-h-Omega-t}
\Omega_{{\rm GW,r}}(k,\tau) 
= {K_h(k,\tau)} \, 
\ncpl_{\rm r}^2 \, \Omega_{t,{\rm r}}(k)\,,
\end{align}
where $\Omega_{t,{\rm r}}(k)\approx c_t^2 {\cal P}_{t,k}(k)/3\Mpl^2$. Thus, apart from the $k$-dependence of $K_h$, the shape of $\Omega_{{\rm GW},{\rm r}}(k)$ 
directly reflects that of $\Omega_{t,{\rm r}}(k)$.
This is a direct consequence of the linearity in our setup,
and is in sharp contrast with the case of the scalar-induced GWs 
where the resultant shape has a rather non-trivial dependence on the original
shape of the curvature perturbation 
power spectrum \cite{Kohri:2018awv,Domenech:2021ztg}. 

The GW density spectrum today is obtained from that during radiation dominance \eqref{Omega-h-Omega-t}. 
For $c_t<1$  \cite{Caprini:2018mtu,Domenech:2021ztg}
\begin{align}\label{GWs-sd-0-v}
\Omega_{{\rm GW},0}(k)
\sim10^{-5}\ncpl_{\rm r}^2 \, \Omega_{t,{\rm r}}(k) \,;~ c_t<1,
\end{align}
where we have assumed ${K_h(k,\tau)}={\cal O}(1)$. 
On the other hand, in the case of $c_t=1$, 
$K_h$ may become very large to cancel the $\ncpl_{\rm r}^2$ factor.
In the limit $\ncpl_{\rm r}\ll1$, we have $K_h=\Delta{\cal N}^2/4$. 
Therefore,
\begin{align}\label{GWs-sd-0-v:ct=1}
\Omega_{{\rm GW},0}(k)
\sim10^{-5} \ncpl_{\rm r}^2\Delta{\cal N}^2 \Omega_{t,{\rm r}}(k) \,;
~ c_t=1,~\ncpl_{\rm r}\lesssim\Delta{\cal N}^{-1}.
\end{align}
We note, however, since $\Delta{\cal N}\lesssim 50$ in the actual universe, $K_h$ cannot be arbitrarily large. For $k$ around the LISA band, $\Delta{\cal N}\sim30$,
we find $K_h=\Delta{\cal N}^2/4\sim220$. 

Another case of interest is $c_t=1$ and $\ncpl_{\rm r}\gtrsim\Delta{\cal N}^{-1}$.
In this case, the sinusoidal oscillation in $K_h$ persists, and we have
\begin{align}\label{GWs-sd-0-v:ct=1_2}
\Omega_{{\rm GW},0}(k)
\sim10^{-5}\sin^2\left[(\ncpl_{\rm r}/2)(\Delta{\cal N}_p+\ln(k/k_p)\right]
\nonumber\\
\times\Omega_{t,{\rm r}}(k) \,;
~ c_t=1,~\ncpl_{\rm r}\gtrsim\Delta{\cal N}^{-1},
\end{align}
where $\Delta{\cal N}_p$ is the number of $e$-folds from the horizon crossing for the mode $k=k_p$.
This leads to two interesting consequences.
First, let us fix $k=k_p$ and vary $\ncpl_{\rm r}$.
Then the sine function is maximum at $\ncpl_{\rm r}=(2n+1)\pi/\Delta{\cal N}_p$ ($n=0,\pm1,\cdots$).
Thus, one obtains a significant enhancement for models with these values of $\ncpl_{\rm r}$. For example, assuming $|\ncpl_{\rm r}|<1$, this happens at
$\ncpl_{\rm r}\sim\pm0.1$, $\pm0.3$, $\pm0.5$, for $\Delta{\cal N}_p\sim30$.
Second, let us fix the coupling to $\ncpl_{\rm r}=2\pi/\Delta{\cal N}_p$. 
Then there appear maxima at $\ln(k/k_p)=(2n+1)\pi/\ncpl_{\rm r}$. 
For $\ncpl_{\rm r}=2\pi/\Delta{\cal N}_p$, this implies the appearance of adjacent
peaks at $k=k_p\exp[\pm2\pi/\ncpl_{\rm r}]$. 
These additional peaks are, unfortunately, physically irrelevant within the applicability of effective field theory, in which one should assume $\ncpl_{\rm r}\ll1$.
Nevertheless, it is intriguing to note that, if we consider the case $\ncpl_{\rm r}\gtrsim1$ from a purely phenomenological point of view, such a model predicts multiple peaks in the GW spectrum.

{\bf Observational constraints.}
Here we consider constraints on $\Omega_{t,{\rm r}}(k)$. 
First, we note that there is a theoretical constraint that the energy density of the additional tensor field should be subdominant during inflation. It is easy to see that this is satisfied if ${\cal P}_t(k,\tau)\ll M_{\rm Pl}^2$, which implies $\Omega_{t,{\rm r}}(k)\ll1$ for all $k$. Moreover, enhanced superhorizon $t_{\bf k}$ modes provide nonlinear source for the linear equation of energy density scalar perturbations at small scales $k\sim{k}_p$. Avoidance of overproduction of PBHs then gives another theoretical bound \cite{Nakama:2015nea}. However, this bound is very weak since $t_{ij}$ is a spectator field \cite{Gorji:2023sil}. Other constraints on the energy density scalar perturbations will also lead to bounds on $\Omega_{t,k}(k)$ at different scales.\footnote{We thank Kazunori Kohri for pointing out this type of bounds.}

Now let us turn to observational constraints.
For scales close to the CMB scale, it seems there exists no apparent constraint.
Since the modes $k\sim k_{CMB}$ come inside the horizon after the universe has become matter-dominated, its contribution to the energy density of the universe seems very small. 
Then the question is how large the induced metric tensor perturbation would be on those scales. 
But the answer depends very much on what we assume for the coupling $\ncpl$.
Depending on the behavior of $\ncpl$ around and after the matter-radiation equality,\footnote{It is also worth mentioning that, assuming $\ncpl_{\rm r}$ does not change significantly after radiation dominance, one can constrain $\ncpl_{\rm r}$ from the distortion of the waveform of $h_{ij}$ \cite{BeltranJimenez:2019xxx}. However, the effect is negligible for $\ncpl_{\rm r}={\cal O}(1)$ \cite{Ezquiaga:2021ler}.}
the value as large as $\Omega_{t,{\rm r}}(k)=O(1)$ could be allowed. 
In any case, since such scales are beyond the scope of the current paper, we leave it for future studies. 

For much smaller scales, $k_p={\cal O}(10^7-10^{18})\mbox{Mpc}^{-1}$ (see Fig.~\ref{fig:1}), the energy density of the tensor perturbation makes an additional
contribution to relativistic degrees of freedom $\Delta{N}_{\rm eff}$ during radiation dominance. This affects the big bang nucleosynthesis as well as the CMB anisotropy. 
As their contribution is expressed as $\Delta{N}_{\rm eff}=\frac{8}{7}\left(\frac{11}{4}\right)^{4/3}\left( \Omega_{{\rm GW,r}}+ \Omega_{{\rm t,r}} \right)$ \cite{Caprini:2018mtu}, using 
$\Omega_{{\rm GW,r}}\leq\Omega_{t,{\rm r}}$,
the maximum contribution is 
$\Delta{N}_{\rm eff,max}=\frac{16}{7}\left(\frac{11}{4}\right)^{4/3}\Omega_{{\rm t,r}}$.
Thus, as a conservative upper bound, we obtain
\begin{align}\label{DeltaN-C}
\Omega_{t,{\rm r}}(k)
= \frac{7}{16} \left(\frac{4}{11}\right)^{4/3} \Delta{N}_{\rm eff,max}
< 0.034\,,
\end{align}
where we have used the current bound $\Delta{N}_{\rm eff}<0.3$ \cite{Planck:2018vyg}. Choosing $\Omega_{t,{\rm r}}\lesssim10^{-2}$ to respect \eqref{DeltaN-C}, we find
\begin{align}\label{GWs-sd-0-UB}
\Omega_{{\rm GW},0}(k) \lesssim
10^{-7}K_h(k,\tau_{\rm m.r.})\ncpl_{\rm r}^2 \,,
\end{align}
where $\tau_{\rm m.r.}$ is the matter-radiation equality time.

As we have previously discussed, we have $K_h={\cal O}(1)$ for $c_t<1$.
Hence, the GW spectrum is suppressed by $\ncpl_{\rm r}\ll1$.
On the other hand, for $c_t=1$, the factor $K_h(k,\tau_f)\ncpl_{\rm r}^2$ can be as large as unity, depending on the scale $k$ and the value of the coupling $\ncpl_{\rm r}$,
even in the case $\ncpl_{\rm r}\ll1$.
In addition, if we regard the action \eqref{action} as a purely phenomenological model, and allow the case of $\ncpl_{\rm r}={\cal O}(1)$, the GW spectrum is not only unsuppressed but may have multiple peaks due to the resonance amplification by forced oscillation.

In any case, the possibility of having a strong GW signal makes our model observationally distinguishable from the usual secondary scalar-induced GWs scenarios. In scalar-induced models, GW signals $\Omega_{{\rm GW},0}(k)\gtrsim10^{-10}$ usually imply the formation of PBHs. Conversely, too large $\Omega_{{\rm GW},0}(k)$ would imply an over production of PBHs. Hence, it cannot be larger than $10^{-10}$. 
On the other hand, in our scenario, we expect signals as large as $\Omega_{{\rm GW},0}\sim10^{-7}$ without producing any PBHs. This is a distinct feature of our scenario.

To be more specific, let us consider the log-normal power spectrum \cite{Pi:2020otn},
\begin{align}\label{PS-Lognormal}
\Omega_{t,{\rm r}}(k) = \frac{A_t}{\sqrt{2 \pi }\Delta_t } \exp\left[-\frac{\ln^2\left(k/k_p\right)}{2 \Delta_t^2}\right] \,,
\end{align}
where $A_t$ is the dimensionless amplitude normalized as $\int_0^\infty\D\ln{k}\,\Omega_{t,{\rm r}}(k)=A_t$ 
and $\Delta_t$ determines the width of the peak around $k=k_p$. It is worth mentioning that we are not restricted to the log-normal power \eqref{PS-Lognormal} and there are many other possibilities. In this regard, \eqref{PS-Lognormal} is not the prediction of our model but instead a possible subset of it. If one considers a subset of action \eqref{action} with explicit functional forms for the couplings, there will be a prediction for the shape of $\Omega_{t,{\rm r}}(k)$.
\begin{figure*}
	\centering
	\begin{minipage}[b]{.4\textwidth}
		\includegraphics[width=1\linewidth]{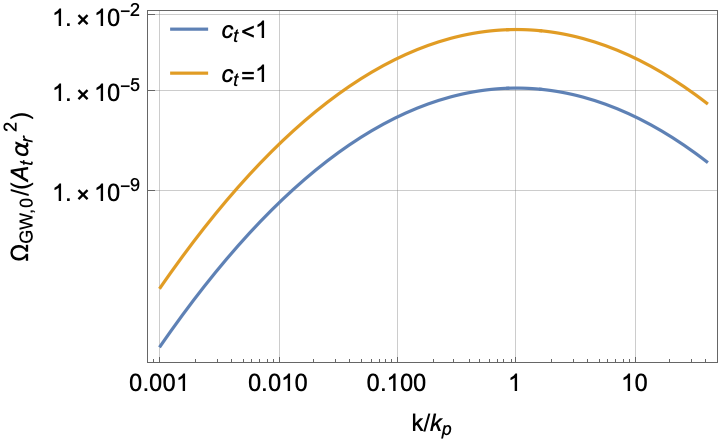}
		\caption{For $\ncpl_{\rm r}\lesssim \Delta{\cal N}^{-1}\ll1$ and $c_t=1$ (see Eq. \eqref{GWs-sd-0-v:ct=1}), an enhancement due to the resonance arises. We have set $\Delta{\cal N}=30$ and $\Delta_t=1$.}\label{fig:2}
	\end{minipage}\qquad
	\begin{minipage}[b]{.4\textwidth}
		\includegraphics[width=1\linewidth]{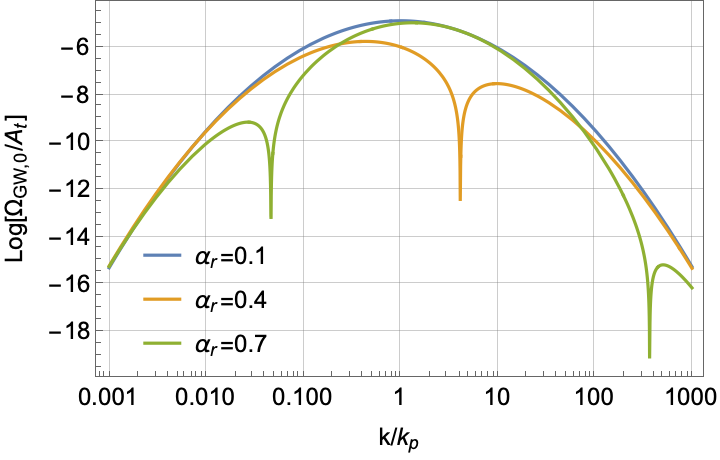}
		\caption{For $\ncpl_{\rm r}\gtrsim \Delta{\cal N}^{-1}\sim1$ and $c_t=1$ (see Eq. \eqref{GWs-sd-0-v:ct=1_2}), multiple peaks show up. We have set $\Delta{\cal N}=30$ and $\Delta_t=1$.}\label{fig:3}
	\end{minipage}
\end{figure*}
In Fig.~\ref{fig:2}, the resulting GW spectra \eqref{GWs-sd-0-v} and \eqref{GWs-sd-0-v:ct=1} for $\ncpl_{\rm r}\lesssim\Delta{\cal N}^{-1}\ll1$ are compared. Clearly, there is an enhancement for $c_t=1$ due to the resonance. Assuming $\ncpl_{\rm r}\gtrsim\Delta{\cal N}^{-1}\sim1$, \eqref{GWs-sd-0-v:ct=1_2} is plotted in Fig.~\ref{fig:3} which leads to oscillatory multiple peaks in the GW spectrum.

Note that, although the amplitude of GW and the peak frequency $k_p$ are free parameters in our effective field theory setup, GW observations restrict their values. For example, non-observation of GW signals in the frequency bands of GW detectors such as LIGO/VIRGO/KAGRA, ET, DECIGO, LISA, PTA, Taiji, and TianQin (see, e.g., \cite{Domenech:2021ztg} for the amplitude and frequency range of GW detectors) imposes bounds on the amplitude and frequency range. Conversely, the observation of a GW signal can be utilized to determine the values of the amplitude and peak frequency in our setup \cite{Gorji:2023sil}.


{\bf Summary.}
While it is usually expected that the contribution of a spectator field to the stochastic GW background is very small compared with the curvature-perturbation-induced GWs, we showed that this is not the case if it provides an extra tensor perturbation on top of the metric tensor perturbation. 
The reason is that an extra tensor perturbation can couple to the metric tensor perturbation at linear order, while the curvature perturbation couples at second order. Implementing the effective field theory method, we considered a general model which captures universal features of an extra tensor perturbation in a model-independent manner. We found a simple analytical expression for the corresponding energy density of the primordial-tensor-induced stochastic GWs. The amplitude and frequency of the produced GWs lie well within the sensitivity ranges of the new generation of GW detectors ET, DECIGO, LISA, and PTA. In particular, if the sound velocities of the two tensor perturbations coincide, the induced GW amplitude may significantly increase due to resonance through forced oscillation, even in the case of a small coupling between them. The mechanism for this oscillation is reminiscent of the well-known process of neutrino oscillation. A distinct feature of our scenario is that, since the tensor perturbation can hardly lead to the formation of PBHs, we expect no presence of PBHs, in contrast to the usual scalar-induced case, in which the detection of strong enough induced GWs suggests the existence of PBHs.
\\


{\bf Acknowledgments.}
This work was supported in part by JSPS KAKENHI Grants Nos. 17H02890, 19H01895, 20H04727, and 20H05853.
The work of MAG. was supported by Mar\'{i}a Zambrano fellowship. We thank Jacopo Fumagalli, Jaume Garriga, Fazlollah Hajkarim, Teruaki Suyama, Valeri Vardanyan, Vicharit Yingcharoenrat, and Ying-li Zhang for useful discussions. MAG. thanks organizers of the workshop ``Non-linear aspects of cosmological gravitational waves" at Kavli Institute for the Physics and Mathematics of the Universe (IPMU) where this work was initiated. 
MAG also  thanks IPMU and Yukawa Institute for Theoretical Physics (YITP) for hospitality and support where this work was in its final stage. 


\bibliography{ref}

\end{document}